# The whack-a-mole governance challenge for AI-enabled synthetic biology: literature review and emerging frameworks

## Corresponding author

Trond Arne Undheim, Ph.D

Stanford Existential Risk Initiative (SERI), Center for International Security and Cooperation (CISAC), Stanford University, 616 Jane Stanford Way, Encina Hall, Room: C240, Stanford, CA 94305, USA

## Abstract

AI-enabled synthetic biology has tremendous potential but also significantly increases biorisks and brings about a new set of dual use concerns. The picture is complicated given the vast innovations envisioned to emerge by combining emerging technologies, as AI-enabled synthetic biology potentially scales up bioengineering into industrial biomanufacturing. However, the literature review indicates that goals such as maintaining a reasonable scope for innovation, or more ambitiously to foster a huge bioeconomy don't necessarily contrast with biosafety, but need to go hand in hand. This paper presents a literature review of the issues and describes emerging frameworks for policy and practice that transverse the options of command-and-control, stewardship, bottom-up, and laissez-faire governance. How to achieve early warning systems that enable prevention and mitigation of future AI-enabled biohazards from the lab, from deliberate misuse, or from the public realm, will constantly need to evolve, and adaptive, interactive approaches should emerge. Although biorisk is subject to an established governance regime, and scientists generally adhere to biosafety protocols, even experimental, but legitimate use by scientists could lead to unexpected developments. Recent advances in chatbots enabled by generative AI have revived fears that advanced biological insight can more easily get into the hands of malignant individuals or organizations. Given these sets of issues, society needs to rethink how AI-enabled synthetic biology should be governed. The suggested way to visualize the challenge at hand is whack-a-mole governance, although the emerging solutions are perhaps not so different either.

## 1 Introduction

Synthetic biology, the multidisciplinary field of biology attempting to understand, modify, redesign, engineer, enhance, or build biological systems with useful purposes (El Karoui, Hoyos-Flight and Fletcher, 2019; Singh *et al.*, 2022; Plante, 2023), has the potential to advance food production, develop new therapies, regulate the environment, generate renewable energy, edit the genome, predict the structure of proteins, and invent effective synthetic biological systems, and more (Yamagata, 2023). It is arguably moving from the lab to the marketplace (Hodgson, Maxon and Alper, 2022; Lin, Bousquette and Loten, 2023). However, the immense promise of synthetic biology has been subject to much hype and it is a paradox that it is still a nascent technology that has not scaled beyond the microscale (Hanson and Lorenzo, 2023). The next major breakthrough might relate to plants (Eslami *et al.*, 2022) or even to mammalian systems (Yan *et al.*, 2023). Despite the small scale, the intermediate term risks are significant, and include contaminating natural resources, aggravation of species with complex gene modifications, threats to species diversity, abuse of biological weapons, laboratory leaks, and man-made mutations, hurting workers, creating antibiotic resistant superbugs, or damaging human, animal, or plant germlines (Hewett *et al.*, 2016; O'Brien and Nelson, 2020; Nelson *et al.*, 2021; Sun *et al.*, 2022). Some even claim synthetic biology produces potential existential risks from lab accidents or engineered pandemics (Ord, 2020), especially in combination with AI (Boyd and Wilson, 2020).

AI-enabled synthetic biology, while surely adding to the risk calculations, has tremendous medium term potential to provide a vehicle for scaling synthetic biology so it may finally deliver on its promise (Hillson *et al.*, 2019; A Dixon, C Curach and Pretorius, 2020; Ebrahimkhani and Levin, 2021; Bongard and Levin, 2023). That being said, the prospect of AI-enabled synthetic biology significantly increases biorisks and particularly brings about a new set of dual use concerns (Grinbaum and Adomaitis, 2023). Although biorisk is subject to an established governance regime (Mampuys and Brom, 2018; Wang and Zhang, 2019), and scientists generally adhere to biosafety protocols if they receive the appropriate training and build a culture of responsibility (Perkins *et al.*, 2019), even experimental, but legitimate use by scientists could lead to unexpected developments (O'Brien and Nelson, 2020). Additionally, recent advances in chatbots enabled by generative AI, technology capable of producing convincing real-world content, including text, code, images, music, and video, based on vast amounts of training data (Feuerriegel et al., 2023), accelerates knowledge mining in biology (Xiao *et al.*, 2023) but has revived fears that advanced biological insight can get into the hands of malignant individuals or organizations (Grinbaum and Adomaitis, 2023). It also further blurs the boundary between our understanding of living and non-living matter (Deplazes and Huppenbauer, 2009). The picture is complicated given the vast innovations envisioned to emerge by combining emerging technologies, as synthetic biology scales up bioengineering turning it into industrial biomanufacturing. Given these sets of issues, society needs to rethink how AI-enabled synthetic biology should be governed.

The research question in this paper is: what are the most important emergent best practices on governing the risks and opportunities of AI-enabled synthetic biology? Relatedly, is stewardship or laissez-faire governance the right approach? How can humanity seek to maintain a reasonable scope for synthetic biology innovation, and integration of its potential into manufacturing, agriculture, health, and other sectors? Do we need additional early warning systems that enable prevention and mitigation of future AI-enabled biohazards from the lab, from deliberate misuse, or from the public realm?

From these questions, the following hypotheses were derived: [1] there is a nascent literature on the impact of AI-enabled synthetic biology, [2] active stewardship is emerging as a best practice on governing the risks and opportunities of AI-enabled synthetic biology, [3] to achieve proper governance, most, if not all AI-development needs to immediately be considered within the Dual Use Research of Concern (DURC) regime, [4] even with the appropriate checks and balances, with AI-enabled synthetic biology, industrial biomanufacturing can conceivably scale up beyond the microscale within a decade or so.

The paper first describes the methods used for the literature review followed by a presentation of the results. A discussion of these findings ensues, addressing the research question and support for the hypotheses, followed by a brief conclusion and suggestions for further research.

## 2 Methods

The purpose of this paper is to conduct a literature review of the issues surrounding AI-enabled synthetic biology and present a set of recommendations for policy and practice. The research goal is to show that a reasonable scope for innovation can be maintained even with instituting early warning systems that enable prevention and mitigation of future AI-enabled biohazards.

Literature review (Sauer and Seuring, 2023), a comprehensive summary of existing research on the topic, was pursued because AI's influx into the synthetic biology field is a very recent development. There is a need to identify gaps in the knowledge that ensue from AI's emerging impact. The goal is to align the AI literature with the synthetic biology risk literature, and develop a theoretical framework for future research. The approach broadly followed Preferred Reporting Items for Systematic Reviews and Meta-Analyses guidelines (PRISMA) for systematic reviews, using formal, repeatable, transparent procedures with separate steps for identification, screening, eligibility, and inclusion of papers (Moher *et al.*, 2015). That being said, in the end a mix of search terms (clearly identified below), plus backward/forward citation searches, where used, which means attempts at replication might yield slightly different results (see Fig. 1). However, because of the nascent research field, in this case, the benefit of flexibility outweighs the costs.

*Figure 1. PRISMA diagram for literature review and citation analysis*

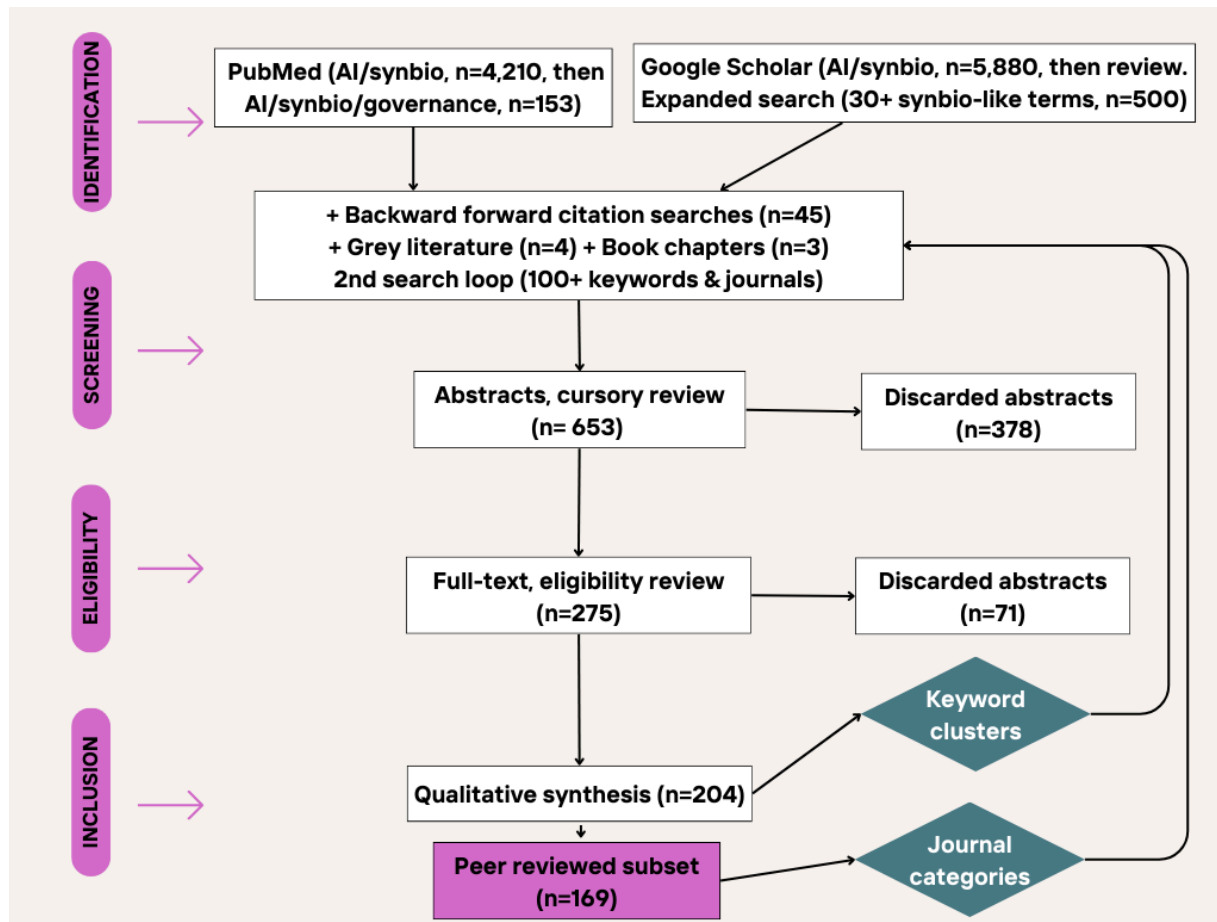

A literature review using the search terms "generative AI", "synthetic biology" and "governance" in Google Scholar generated only 97 results, so the search was broadened to "AI", which generated 5,880 results, also capturing important articles before the generative AI discoveries of 2022-2023. Similar searches in Scopus (artificial AND intelligence AND synthetic AND biology AND governance) generated only 9 documents. Using the terms "AI synthetic biology governance" in PubMed generated only 14 results of which only 1 paper (on synthetic yeast research and techno-political trends) was retained. However, removing the term governance gave 4,210 results. The search was limited to 2020-2024 publication dates, and further filtered to only review or systematic reviews to get 153 results which were screened down to 14 relevant articles.

Searching Social Sciences Citation Index (pub dates 2021-2023) searches yielded 257 results for "synthetic biology", which were all reviewed, and 13 abstracts were selected into the sample. Searching Business Source Complete (pub dates 2020-2023) for 'synthetic biology' yielded 183 academic journal papers, 15 of which were relevant and from which 7 were retained after deduplication (this search was finalized last). Other search terms such as 'bioeconomy' and 'governance' performed better in this database.

The final search protocol borrows from Shapira et al. who note that papers that don't explicitly use "synthetic biology" in their title, abstract or key words could still be relevant because it is an interdisciplinary field (Shapira, Kwon and Youtie, 2017). Shapira et al. track the emergence of synthetic biology over the 2000-2015 period, first retrieving benchmark records, extract keywords from there, and then searching. With that insight, having selected 150 articles, read their abstracts, indexed their keywords, and scanned the content of all papers, I then went back, did new searches based on the keyword clusters that seemed promising, and, as a result, found additional papers to include in the sample.

The analysis was also complemented with papers that did discuss the overall impact of generative AI on biology or science, or research, using the search terms: "generative AI" AND "Science" OR "research". Because generative AI is such a recent term, the search was expanded to preprints in the gray literature. The final research protocol included a much wider set of search terms, including a fuller set of keywords such as AI risk, bioethics, bioinformatics, biohacking, biorisk, biosafety, biosecurity, computational biology, DIY biology, Do-It-yourself laboratories, dual risk, dual-use research of concern (DURC), emerging technology, generative AI, industry, large language models (LLMs), multi-omics, risk mitigation, systems biology, AI-bio capabilities, chatGPT, biomanufacturing, biosurveillance, bioweapons (always used in combination with AI and/or risk). Separate searches for 'synthetic biology' AND legislation OR 'policy' OR 'regulation' were also conducted. Similarly, when few papers were found on the management and industry aspects, specific searches on 'synthetic biology' AND/OR 'startups', 'industry', 'market', and 'economy' were pursued.

The final inclusion criteria involved any type of published scientific research or preprints (article, review, communication, editorial, opinion, etc.) as well as any high quality article (based on subjective review) published by a government agency, think tank or consulting firm. A total of 653 abstracts were considered, but only 204 sources and a subset of 169 peer reviewed papers were included in the final review (see [Appendix A-papers in sample](#)), representing 111 different journals (average impact factor: 9.94) from 4 fields. The overwhelming number of papers (114) originated from journals in Science, Engineering & Technology, 36 from interdisciplinary journals and only 9 from Social science and Humanities journals and 9 from Management journals (see [Appendix B-journals in sample](#)), as well as 4 preprints and 3 other types of publications (such as chapters in books as well as think tank white papers and memos). Once papers were identified for synthesis and analysis, 6-10 keywords were manually extracted from each article, starting with the ones identified by the authors (if any), and the diversity of journal types was recorded.

No human data was collected for this study. However, ethical considerations, such as how to discuss whether synthetic biology is significantly different from nature, were carefully addressed throughout the study.

The study's findings may not be generalizable to biological research that does not rely on synthetic approaches or that only have limited use of AI technologies. Given that scale-up seems to be a much desired future development that industry and researchers both expect, future research could explore the complex factors influencing the scale-up of industrial biomanufacturing.

## 3 Results

There was no significant concentration of papers in any specific journal, instead the topic was covered broadly across journals. However, 23 percent of the journals (25 journals) were published by Elsevier, and 23 percent of the journals (25 journals) were published by *Springer Nature,* each highly overrepresented in the sample. The world's two top publishers (in number of published journals) each publish nearly 3000 journals (Curcic, 2023).) The showing of the third (Taylor & Francis, 2508 journals total 8 in our sample), and fourth (Wiley, 1,607 journals total, 6 in our sample) was far lower, grouped with the fifth (Oxford Academic, 7 journals in sample), and sixth (MDPI, 5 journals in sample), who only publish about 500 journals total (Curcic, 2023). The country of publication provided another slight surprise compared to data presented by Shapira et al. 's (Shapira, Kwon and Youtie, 2017) findings of a US and UK dominance when tracking the emergence of synthetic biology over the 2000-2015 period. Our data, in contrast, shows the US a bit behind in synthetic biology publishing (the Netherlands 23 percent, UK 21 percent, US 19 percent, Germany 12 percent, and Switzerland 11 percent). One explanation might be that in several cases US professional societies use a European publisher.

At least 8 breakout papers contained especially innovative, useful, or surprising observations for scholars and policy makers alike (Camacho *et al.*, 2018; Trump *et al.*, 2019; Hagendorff, 2021; Eslami *et al.*, 2022; Hanson and Lorenzo, 2023; Holzinger *et al.*, 2023; Sundaram, Ajioka and Molloy, 2023; Yan *et al.*, 2023), each summarized in a sentence:

[1] As long as the black box issues of deep learning models are addressed they will transform insights into molecular components and synthetic genetic circuits and reveal the design principles behind so one can iterate rapidly and create complex biomedical applications (Camacho *et al.*, 2018).
[2] An interdisciplinary approach between the physical and social sciences is necessary (and seems to be proceeding), fostering sustainable, risk-informed, and societally beneficial technological advances that are driven by safety-by-design and adaptive governance that properly reflects uncertainty (Trump *et al.*, 2019).
[3] Machine learning for synthetic biology can yield (forbidden) knowledge with dual-use implications that needs to be governed given legitimate misuse concerns that we have seen in other areas such as nuclear energy (Hagendorff, 2021).
[4] If synthetic biology can deploy the Design-Build-Test-Learn (DBTL) cycle, bridging the cultures of bench scientists and computational scientists, and properly quantify uncertainty, it will impact every activity sector in the world (Eslami *et al.*, 2022).
[5] For the field of synthetic biology, considering all the hype, it is high time to deliver, likely by toning down claims to have all the answers and capitalize on the achievable goals, and enlist tool builders in universities, realizing that biofoundries will not be generalized industrial factories near term but will remain fermentation plants for enzymes (Hanson and Lorenzo, 2023).
[6] AI is already ubiquitous in biotechnology (Holzinger *et al.*, 2023).

[7] In the future, AI will be a driving force of biotechnology whether we like it or not (Sundaram, Ajioka and Molloy, 2023).
[8] So far, AI in synthetic biology has been used for foresight, data collection, and analysis but in the future it will be used to design complicated systems (Yan *et al.*, 2023).

The 1297 unique keywords found in these 204 sources were clustered into 81 broad categories, 42 of which seemed particularly important as literature search keywords (see Fig. 2).

*Figure 2. Keyword categories relevant to Synthetic biology, AI, governance and innovation*

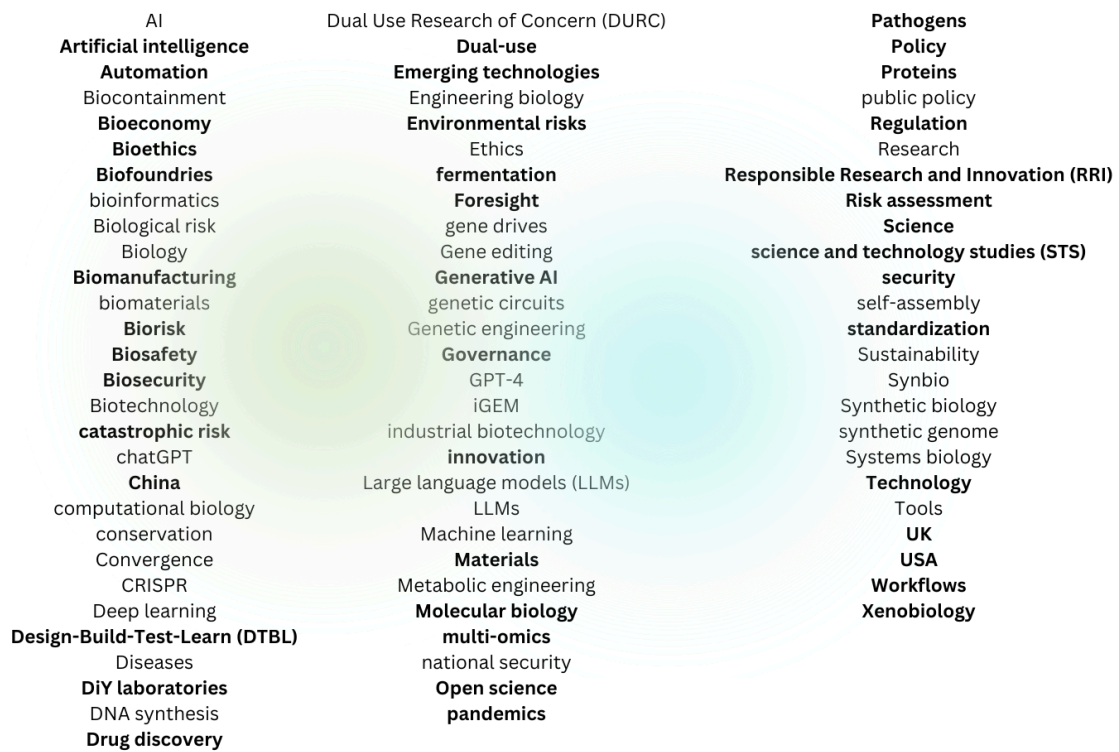

These categories were further reduced into 8 clusters representing key issues: *Applications* (drug discovery), *Bioeconomy* (biomanufacturing, innovation), *Countries* (China, EU, UK, US), *Governance* (bioethics, biosafety, dual use, risk assessment), *Science* (computational biology, Design-Build-Test-Learn (DTBL), materials, molecular biology, open science, RRI, STS, xenobiology), *Tools* (artificial Intelligence, DIY laboratories, fermentation, generative AI, biofoundries, emerging technologies, laboratories, multi-omics, technologies, xenobiology, workflows), *Materials* (genes, proteins), and *Risks* (biological, environmental, pandemics) (see Fig. 3).

*Figure 3. Synbio issue clusters*

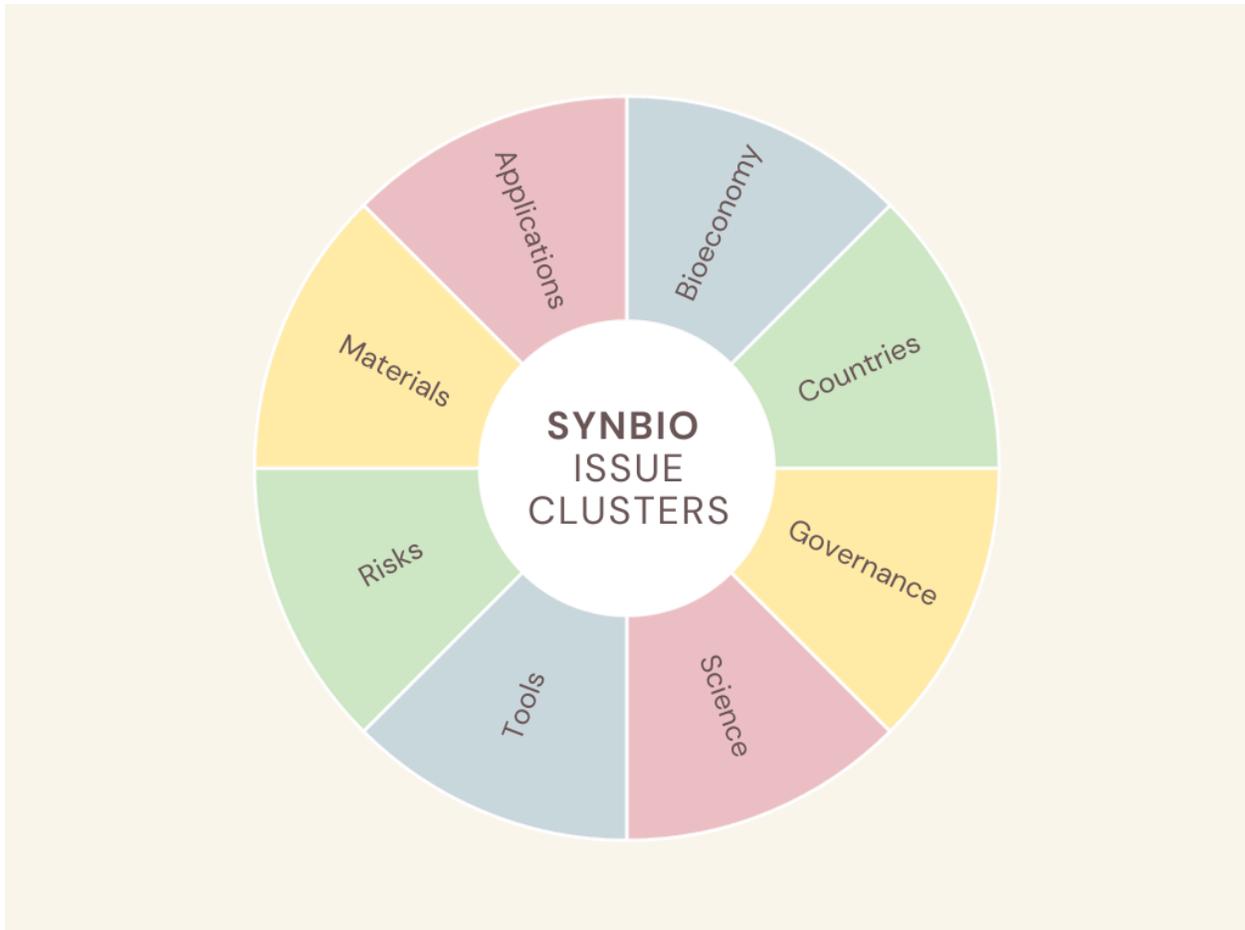

The presentation of results is organized according to the initial four hypotheses on : [1] the nascent literature, [2] best practices, [3] the DURC regime, and [4] scale up.

## 3.1 The impact of AI-enabled synthetic biology

Among the 169 peer-reviewed papers in the sample, there were 81 papers that explicitly discussed the impact of AI-enabled synthetic biology. The other 88 papers discussed risk but not explicitly from AI. Equally surprising was the exclusion of AI in all the other papers, given that several were review articles or otherwise covered state-of-the art or emerging technologies and tools for synthetic biology. There is no ready explanation for this omission, except to say that perhaps (a) those researchers are not familiar with the potential of (generative) AI for biology, (b) don't think it is as big of a deal as others do, or (3) consider it less relevant for today's concerns in synthetic biology, or (4) consider AI (machine learning) an essential tool but prefer not to elevate it beyond its obvious place as a key research tool.

Another finding is that many papers that I found relevant to the future of synthetic biology's governance, risk, and innovation trajectory, did not in fact use that term. Dozens and dozens of papers included in the sample happily discussed AI and the impact on their field, be it metagenomics of the microbiome (Wani *et al.*, 2022), health and intelligent medicine (Achim and

Zhang, 2022), applied microbiology (Xu *et al.*, 2022), designer genes (Hoffmann, 2023), drug discovery (Yu, Wang and Zheng, 2022), oncology (Wu *et al.*, 2022), systems biology (Helmy, Smith and Selvarajoo, 2020) without realizing that synthetic biology is bound to intersect with it at some point soon (or at least explicitly omitting the use of the term). What could the reason be? Is the term upsetting to part of the biology community or establishment? Synbio scholars clearly frame their problems differently from the biology establishment. Perhaps there is a disparity in age, experience, and skills between patchy biological knowledge of bio-IT nerds and lacking IT skills among biologists? AI has been applied to material discovery, finding 700+ new materials so far (Merchant *et al.*, 2023) and it is a question of time before it will be used for scalable materials design using AI-enabled synthetic biology (Tang *et al.*, 2020; Burgos-Morales *et al.*, 2021) although for real world applications we might first need better standardized vocabularies for biocompatibility (Mateu-Sanz *et al.*, 2023).

Only 5 papers (Kather *et al.*, 2022; Grinbaum and Adomaitis, 2023; Morris, 2023; Ray, 2023; Xiao *et al.*, 2023), a popular science article (Tarasava, 2023), and an editorial ('Generating "smarter" biotechnology', 2023) discussed the impact of generative AI on synthetic biology. This is expected to increase dramatically quite soon, given the success of this latest wave of AI technology and the platform aspects of its spread. However, as one paper put it, synthetic biology has a natural synergy with deep learning (Beardall, Stan and Dunlop, 2022). The use cases discussed in various papers include: as a classifying text, generic search engine, generating ideas, helping with access to scientific knowledge, coding, patient care (Clusmann *et al.*, 2023), protein folding, proofreading, sequence analysis, summarizing knowledge, text mining of biomedical data, translation (Clusmann *et al.*, 2023), workflow optimization, foresight of future research directions (Yan *et al.*, 2023); collection of related synthetic biology data, and more (Beardall, Stan and Dunlop, 2022; Clusmann *et al.*, 2023; Tarasava, 2023). The promise of AI-enabled cell-free synbio systems (Lee and Kim, 2023), which use molecular machinery extracted from cells, is particularly significant for automation and scale-up of biosensors among other things. It bears pointing out that the significant advances in cell-free synbio systems enabling the acceleration of biotechnology development, specifically its ability to enable rapid prototyping as well as the ability to conduct predictive modeling, pre-date generative AI by a decade (Moore *et al.*, 2018; Müller, Siemann-Herzberg and Takors, 2020). Even today, he the barriers seem to be limited availability of relevant data either because it does not exist yet, because data is scarce, the data set is small, because it is not publicly available, or because it is not formatted in useful ways (Rosenbush, 2023). Not all of these challenges can be immediately resolved by generative AI.

What matters most to the governance and innovation concern would be those barriers, areas, or workflows where AI could make the biggest impact, not just for the field applying it but for the shared resource that is AI-enabled synbio that would grow the pie. Based on the literature review, I've attempted to suggest which topics fit in that perspective (see Fig 4). For example, progress on interoperability would benefit all, as would AI-enabled lab operations workflows. Each would be a synbio building block. Big ticket items such as protein folding is in a bit of a different category. When AlphaFold achieved near-perfect protein fold predictions, it was the most important moment for AI in science so far, yet left plenty of work for structural biology

(Perrakis and Sixma, 2021), including the application of coiled coils as a self-assembly building block in synthetic biology (Woolfson, 2023). Similarly, when the mRNA platform became a successful vehicle for a COVID-19 vaccine that changed the world, this happened *in vitro*, yet, the production of synthetic mRNA (Hınçer *et al.*, 2023) or miRNA (Matsuyama and Suzuki, 2019) in the cell itself would be an even more significant breakthrough–and getting there might require the use of AI (Naderi Yeganeh *et al.*, 2023).

*Figure 4. AI-synbio accelerants and use cases*

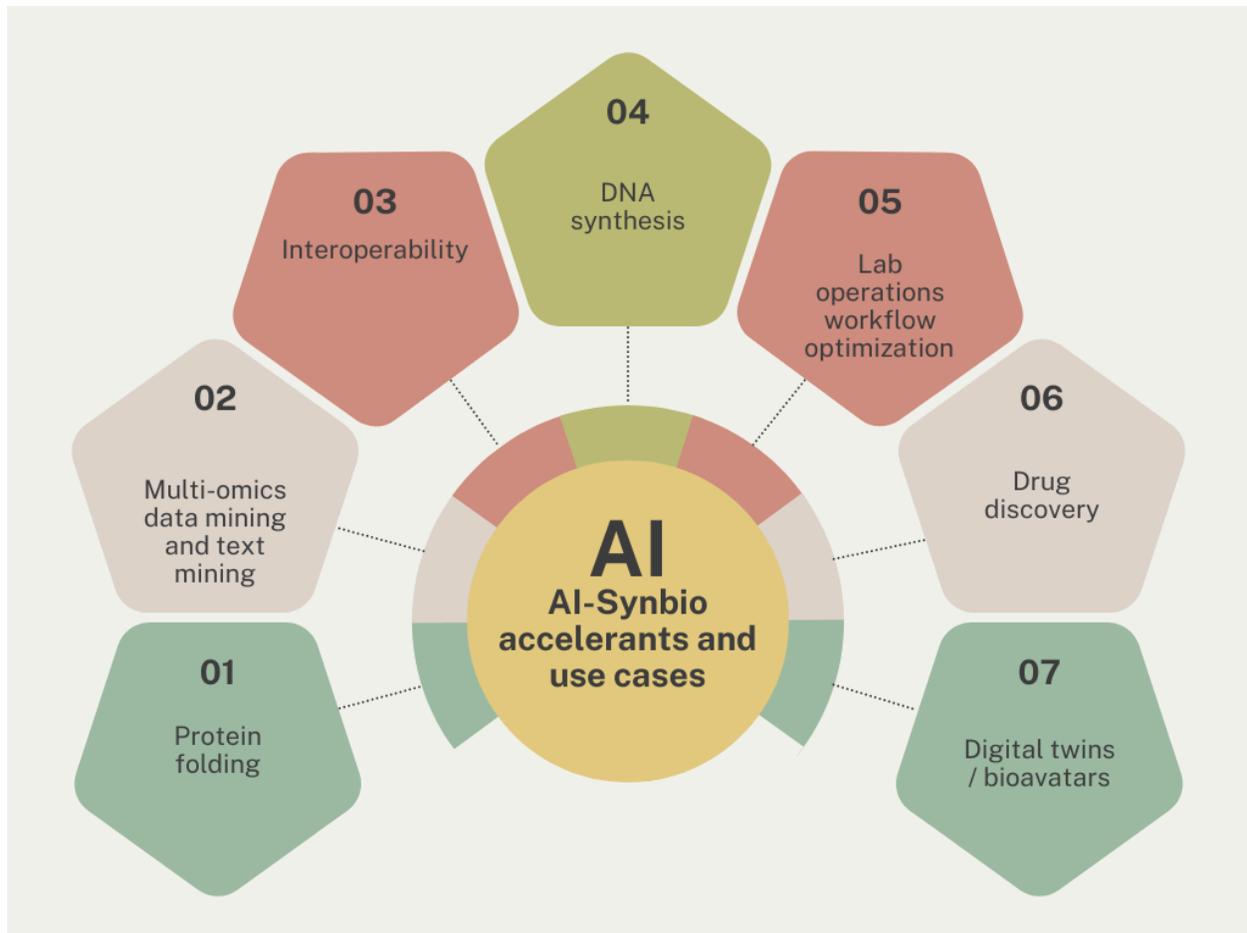

On the other hand, there is no reason to believe that the top labs will be overtaken, quite the contrary, in fields such as consulting, it seems that generative AI accelerates the work of top teams (Moran, 2023). However, the worrying aspect is that the previous assumption from biorisk work pre-generative AI was that developing pathogens is an activity only possible in highly advanced biolabs. The increasing availability of insight, instructions, as well as wetlabs and foundries on demand (Sandberg and Nelson, 2020), would seem to be a potential issue to watch.

## 3.2 Best practice in AI-synbio governance

In the literature, there is ample evidence of what constitutes biosafety and biosafety governance best practice (Perkins *et al.*, 2019; Wang and Zhang, 2019; Li *et al.*, 2021; Mökander *et al.*, 2022; Sandbrink, 2023b) and the emphasis is on a mix of specific training and, relatedly, developing a safety and responsibility work culture. In previous decades, the few advanced labs that existed were "compliant" biocontainment actors, for which acceptable systems were in place. However, regulating synthetic DNA comes with new challenges, including scalability, and less ability to create genetic firewalls to natural organisms, and it has become easier to circumvent oversight (Hoffmann *et al.*, 2023). The more accessible (Wang and Zhang, 2019)and generally useful synbio potentially is regarded to be (Sun *et al.*, 2022), the less likely it is that prohibition will remain an effective tool. Decades-old bioinformatics resources originally developed to compare gene sequences, such as BLAST, have been re-used, with mixed results, as biosafety tools to identify pathogens (Beal, Clore and Manthey, 2023). Newer tools, such as machine learning-based topic models, enable spotting trends across a wide set of biosafety research publications (Guan *et al.*, 2022). AI-synbio governance (Achim and Zhang, 2022; Mökander *et al.*, 2022; Grinbaum and Adomaitis, 2023; Holland *et al.*, 2024) is expected to be more of the above, but also requires AI skills and perspectives that go far beyond wet lab practices and will require updates to biosafety laws, regulation, governance, standardization (Pei, Garfinkel and Schmidt, 2022). It will change the role of the state (Djeffal, Siewert and Wurster, 2022) as it will no longer be the primary norm setter or enforcer of responsibility.

AI is already contributing to the fragmentation of biology (Hassoun *et al.*, 2022) and will challenge medical expertise among specialists (Patel *et al.*, 2009). Generative AI, and especially other advancements in multi-modal AI, combined with better multi-omics synbio dataset interoperability (Topol, 2019) and standardization, will eventually lead to fundamentally new playing fields. Vigilance is required (Harrer, 2023) both to get us there, predict when we will get there, and decide what to do when we get there. The initial issue surrounds AI-synbio lab safety practices (D'Alessandro, Lloyd and Sharadin, 2023) when the "lab" suddenly is a dispersed concept, and decisions around forbidden knowledge (Hagendorff, 2021), new sets of responsibilities in the research community (Blok and von Schomberg, 2023) and among health practitioners (Achim and Zhang, 2022), avoidance of doom speak (Bray, 2023), handling the reality of malicious actors (Carter *et al.*, 2023), and will represent an enormous challenge for reskilling and upskilling those who want to work with the topic (Xu *et al.*, 2022).

Getting it right will mean balancing brave investments (Hodgson, Maxon and Alper, 2022) with monitoring the effects, including developing an ethics and a taxonomy for working with AI-synbio-human hybrids and intelligence (Nesbeth *et al.*, 2016; Damiano and Stano, 2023), dealing with new synthetic pathogens (O'Brien and Nelson, 2020), saying carefully goodbye to the natural world (Lawrence, 2019; Webster-Wood *et al.*, 2022; Bongard and Levin, 2023) or at least radically enhancing biocontainment (Schmidt and de Lorenzo, 2016; Aparicio, 2021; Vidiella and Solé, 2022; Hoffmann, 2023), as well as developing new approaches to worker safety (Murashov, Howard and Schulte, 2020). This leads into the issue of dual use of concern, which currently is a binary issue even though it is about to become immensely complex, requiring a more nuanced approach (Evans, 2022; Sandbrink, 2023a), given the legitimate concern with deliberate, perhaps even deliberate synthetic pandemics (Sandbrink, 2023b).

## 3.3 Broadening the DURC regime

Dual use is mentioned by several papers (Getz and Dellaire, 2018; Torres, 2018; DiEuliis *et al.*, 2019; Alexander Hamilton *et al.*, 2021; Hagendorff, 2021; Grinbaum and Adomaitis, 2023; Vaseashta, 2023). However, it can have broader meaning, for example positively referring to open source (Esquivel-Sada, 2022) as opposed to negatively referring to non-conformant use. The idea of broadening the dual use research of concern (DURC) regime, which gained steam during COVID-19 (Sandbrink, Musunuri and Monrad, 2023) is mentioned in the preprint literature (Grinbaum and Adomaitis, 2023), and alarms about dual-use involving AI are sounded in several papers (Urbina *et al.*, 2022; D'Alessandro, Lloyd and Sharadin, 2023) and in Sandbrink's Ph.D thesis (Sandbrink, 2023b).

Giving a complete regulatory overview of synthetic biology is complex (Beeckman and Rüdelsheim, 2020) and is not the task of this paper, but Table 1 still lists some key standards and guidelines discussed in the sample, and relevant to DURC issues.

*Table 1 List of biorisk standards and guidelines*

Global:
World Health Organization's (WHO) Laboratory Biosafety Manual (2004)
International Health Regulations (WHO 2005)
Biosafety in Microbiological and Biomedical Laboratories (BMBL), 6th ed. (Centres for Disease Control and Prevention (1984-2021) (BMBL, 2023)
Convention on Biological Diversity (CBD) and Protocols (Cartagena & Nagoya)
Biological and Toxin Weapons Convention (BTWC)
Tianjin Biosecurity Guidelines for the Code of Conduct for Scientists (2015)
World Economic Forum Global Future Council on Synthetic Biology

US:
International Compilation of Human Research Standards (ICHRS)
Laboratory Biorisk Management (CWA 15793)
Screening Framework Guidance for Providers of Synthetic Double-stranded DNA (U.S. HHS)
NIH Guidelines for Research Involving Recombinant or Synthetic Nucleic Acid Molecules
Executive Order 14081 to enable the progress of biomanufacturing and biotechnology (2022)
U.S. Executive Order on Safe, Secure, and Trustworthy Artificial Intelligence (2023)

EU:
Advanced Therapy Medicinal Products Regulation (ATMP) 2007

## 3.4 Scaling industrial (bio)manufacturing

Synbio is not yet a mature engineering industry with well-understood costs and timelines (Watson, 2023) and investments fluctuate from year to year (SynBioBeta, 2023). A recent Schmidt futures report defines commercial production scale as a fermentation capacity of 100,000 liters or more and states only a few U.S. companies currently have infrastructure at this scale and relatively inaccessible to small- and medium enterprises at the present moment (Hodgson, Maxon and Alper, 2022). Achieving pilot scale is the first hurdle to pass and would cost in excess of $1 billion to build a dozen pilot facilities to fuel the U.S infrastructure alone (Hodgson, Maxon and Alper, 2022). Synbio was not truly part of the industry 4.0 paradigm either (Jan *et al.*, 2023). The keywords to describe the industrial aspect of synthetic biology included: 'bioeconomy', 'bio-capitalism', 'biomanufacturing', 'biotech industry'. Surprisingly few path breaking peer reviewed articles were found on these topics. The six key ingredients for biomanufacturing derived from our sample are: biological insights, AI, bioprocessing, engineering scale-up, governance frameworks, and gigascale investments (see Fig. 5).

*Figure 5 Biomanufacturing ingredients*

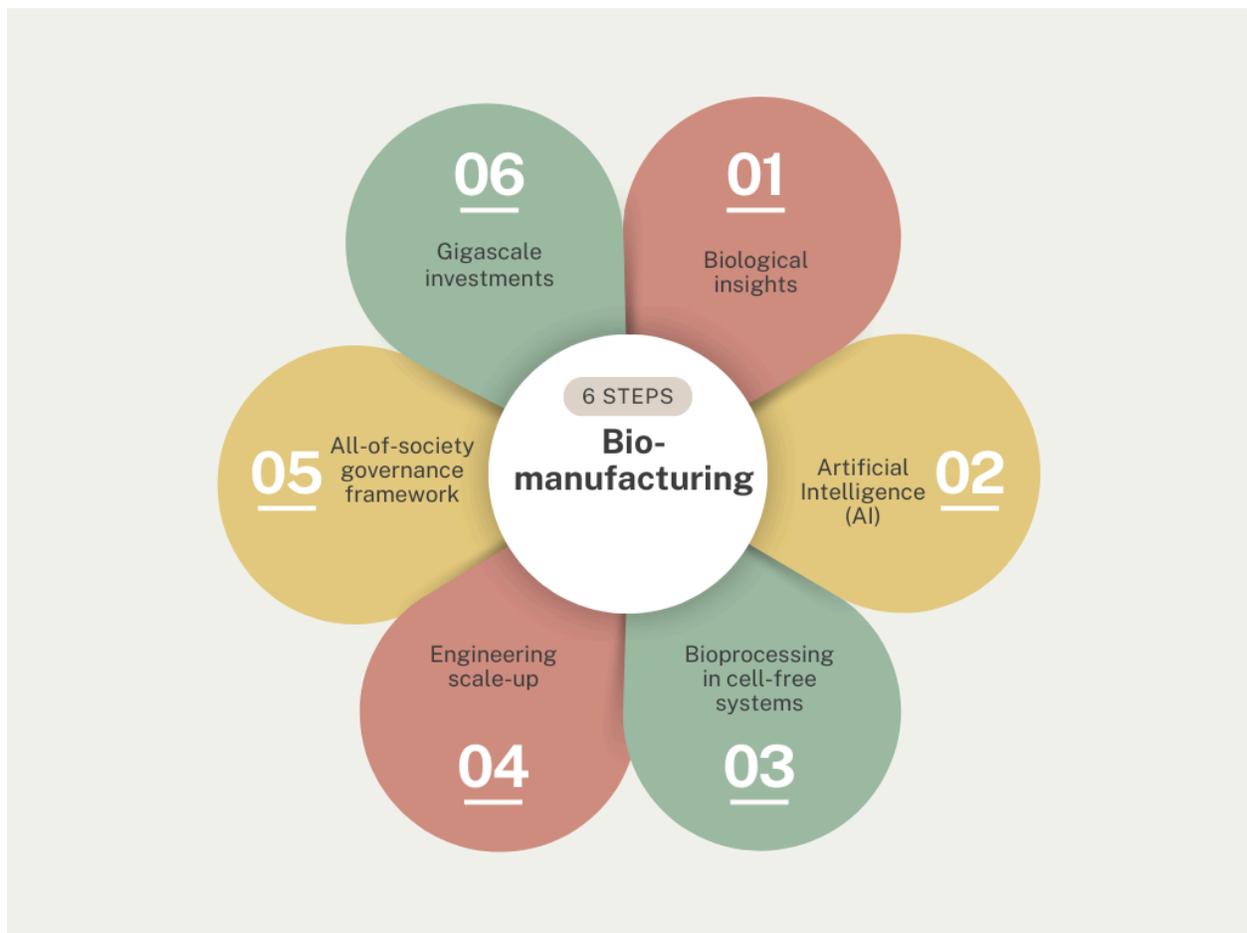

Attempting to pinpoint exactly when a sci-tech paradigm will take off commercially is a fool's errand. Exceptional growth in research communities can be tentatively forecasted from citation analysis (Klavans, Boyack and Murdick, 2020). The emergence of new industries is significantly more complex but the growth in intangible assets (Börner *et al.*, 2018), such as generative AI, applied to an industry (manufacturing) would be a clear indicator. One article in our sample proposed a taxonomy of four innovation types specific to the bioeconomy: Substitute Products, New (bio-based) Processes, New (bio-based) Products, and New Behavior, each carries their own commercialization challenges (Bröring, Laibach and Wustmans, 2020). Deriving insights from other papers, existing or emerging business models in synthetic biology would include automation, contract research, increasing crop yields in agriculture (Bhardwaj, Kishore and Pandey, 2022; Wang, Zang and Zhou, 2022), data driven design (Freemont, 2019), efficiencies, new components, DNA synthesis (Seydel, 2023), infrastructure, licensing, manufacturing molecules for the food industry (Helmy, Smith and Selvarajoo, 2020), modularity, new materials, new platforms, new products, open source tools, services, or substitution, such as a new technology stack (Freemont, 2019).

That being said, despite the relatively low number of papers describing the synbio industry (van Doren *et al.*, 2022), there are signs in the gray literature and in the consulting literature (Candelon *et al.*, 2022) that things are changing within this decade. Arguably, the synbio startup boom in pharma and food industries will be duplicated in health and beauty, medical devices, and electronics, with cost-based competition from syn-bio alternatives in chemicals, textiles, fashion, and water industries, soon to be followed by the mining, electricity, and construction sectors (Candelon *et al.*, 2022). The way it might happen is not necessarily only through flashy, radical innovations, but incrementally because synbio is becoming a useful tool to improve performance, quality, and sustainability of almost all types of manufacturing (Candelon *et al.*, 2022).

Made-to-order synthetic DNA is faster and cheaper than before but is still a massive bottleneck to building scalable biological systems based on synthetic components (Seydel, 2023). The future role of synthetic biology in carbon sequestration into biocommodities could be of major industrial importance provided the bioproduct could be commercialized (Jatain *et al.*, 2021).

# 4 Discussion

One of the papers in the sample reports that synbio discourse is framed in six major ways: as science, social progress, risks and control, ethics, economics, and governance (Bauer and Bogner, 2020), which roughly matches the eight clusters identified based on the papers in the present sample: Applications, Bioeconomy, Countries, Governance, Science, Tools, Materials, and Risks. These frames tend to belong to different camps (particularly citizens, corporations, governments, nonprofits, and startups), with separate agendas and concerns, as opposed to characterizing aspects of a discussion that all actors should be having. There are signs this is changing towards more adaptive approaches to address the uncertainty surrounding the effects of novel technologies (Millett *et al.*, 2020; Mourby *et al.*, 2022) in parts of the system, such as in innovation communities such as iGEM (Millett *et al.*, 2020; Kirksey, 2021; Millett and Alexanian, 2021; Vinke, Rais and Millett, 2022), or in entrepreneurial ecosystems (Nylund *et al.*, 2022). However, those are not characteristic of the governance system as a whole.

## 4.1. Terminological and sectoral confusion, growing pains

Given the nascent state of AI-enabled synthetic biology, there is an overload of related and relatable search terms and keywords that proliferate in the scientific community and online, making it difficult to compare, find, and cluster case studies, research, and policy relevant insight. Even after considerable search efforts, we were left with 1297 unique keywords, which were boiled down to 81 broad categories, and further to 42 literature search keywords. The situation will persist, and in all likelihood, it will get worse before it gets better. There are those hoping for a taxonomic renaissance (Bik, 2017) to remedy the problem, including a taxonomy for engineered living materials (Lantada, Korvink and Islam, 2022). An early article attempted to do the same for the field of synthetic biology (Deplazes, 2009), but it might have been too early in the cycle.

Historically, synbio has been seen as a disruptive innovation yielding products and processes which may not be well aligned with existing business models, value chains and governance systems (Banda and Huzair, 2021), but this might now be changing and synbio approaches get integrated into traditional industries and sectors. That is exciting for industrial innovation but challenging for governance, risk and regulation.

Even though commercially available synthetic biology-derived products are already on the market that are, arguably 'changing the world' (Voigt, 2020), the economics of synthetic biology (Henkel and Maurer, 2007), the biomanufacturing industry overall, is in its infancy. McKinsey might be right that it is a $4 trillion gold rush waiting to happen (Cumbers, 2020), or as BCG claims, $30 trillion by the end of the decade (Candelon *et al.*, 2022), across food and ag, consumer products and services, materials and energy production, and human health and performance (Ang, 2022; Clay and September, 2023). However, the conspicuous absence of management and business articles on synbio in our sample might indicate that the business dimension is so embryonic that these visions are not yet a story worthy of sustained business school attention. The umbrella term bioeconomy (Baker, 2017; Bröring, Laibach and Wustmans, 2020; Marvik and Philp, 2020; Banda and Huzair, 2021; Hodgson, Maxon and Alper, 2022; Bröring and Thybussek, 2023; Clay and September, 2023; Rennings, Burgsmüller and Bröring, 2023) is perhaps convenient, but encompasses so much that it is hard to know what it means..

## 4.2. Transdisciplinary barriers to growth

What seems to be missing in the literature is a clear vision for how AI-enabled synthetic biology would be truly different from previous approaches. Most of the papers imply that AI will remain only one of many technologies relevant to progress in the synthetic biology field. No papers paint a picture where there is a straightforward path to massive scale-up, with possible exception of AI for multi-omics. The shift would happen once the synbio field was able to shift from its current systems-centric approaches (requiring slow, cumbersome wet lab experiments and trial-and-error tinkering) to data-centric bioprocessing approaches (not just using AI for data processing) that are themselves digitally scalable (Owczarek, 2021; Scheper *et al.*, 2021), and constitute automated design-build-test systems (Holland *et al.*, 2024), supported by digital twins

(Manzano and Whitford, 2023). Having said that, enormous efficiencies could be had through even much simpler process automation and operations improvements in biomanufacturing, for example through no-code methods (Linder and Undheim, 2022).

The barriers to the field of synthetic biology's growth are many, from (1) technical feasibility, including the scientific problems connected with the fusion of three disciplines; synthetic biology, artificial intelligence, and social science (Trump *et al.*, 2019), via (2) various forms of risk to (3) industrial challenges, to (4) institutional challenges, or (5) social dynamics.

On the technical side, we find the challenges surrounding data quality (Patel *et al.*, 2009) the fragmentation of knowledge (Hassoun *et al.*, 2022)interoperability (Mateu-Sanz *et al.*, 2023)or standardization (Endy, 2005; Hanczyc, 2020; Garner, 2021; Pei, Garfinkel and Schmidt, 2022; Mateu-Sanz *et al.*, 2023). For example, even though there is great need, and the desire is there, standardizing complex biological systems is difficult (Garner, 2021). As many of the papers in the sample point out, there are also scientific problems connected with the fusion of two disciplines, synthetic biology and artificial intelligence. A multi-layer technology stack is evolving (Freemont, 2019). There is transdisciplinary training and perspective required (Hammang, 2023). There is also considerable uncertainty produced when three domains (or more), and their methodologies, technologies, and tools, are merging (Trump *et al.*, 2019).

Notably, (1) biological insight is needed to deploy AI correctly, yet cell behavior is unpredictable (Lawson *et al.*, 2021)(2) AI insight is needed to capture what ends up retranslated as biological patterns in the data, but AI insight alone is not sufficient to identify what data might be relevant and (3) social science insight, including business models, sociotechnical issues (Marris and Calvert, 2020), social dynamics, social implications, governance, risk, ethics, and psychological reactions, is needed to assess the feasibility of R&D, product development, and commercialization of the emergent field's output. That's a tall order for individual researchers, teams, companies, and political institutions alike. As the field grows in importance, scale, and impact, it will entail an enormous societal reskilling effort (Hammang, 2023).

Industrial challenges include: biosafety (Pei, Garfinkel and Schmidt, 2022), the availability of capital (Baker, 2017; Helmy, Smith and Selvarajoo, 2020; Hodgson, Maxon and Alper, 2022; Sargent *et al.*, 2022), the lack of a scalable manufacturing workflow (Hillson *et al.*, 2019; Ataii *et al.*, 2023), regulatory uncertainty (Huzair, 2021), innovation challenges (Banda and Huzair, 2021; Tait and Wield, 2021), intellectual property rights (Esquivel-Sada, 2022), investment risks (Hodgson, Maxon and Alper, 2022), or worker safety (Murashov, Howard and Schulte, 2020).

Various forms of risk will impact synbio growth, notably AI risk (O'Brien and Nelson, 2020; Grinbaum and Adomaitis, 2023), the potential for a slew of catastrophic risks (DiEuliis *et al.*, 2019) such as new pathogens, or even the specter of existential risks (Boyd and Wilson, 2020) threatening the flourishing or survival of humanity.

On the institutional side, we cannot discount biosecurity (Millett *et al.*, 2020), bioterrorism (Trump *et al.*, 2020; Sheahan and Wieden, 2021; Bray, 2023), the constant challenge of existing, emerging, or evolving bioweapons (Gronvall, 2018; DiEuliis *et al.*, 2019; Trump *et al.*, 2020), the cost of deregulation (Sargent *et al.*, 2022), dual use (Getz and Dellaire, 2018; Ienca and Vayena, 2018; Evans, 2022; Grinbaum and Adomaitis, 2023), global governance (Linkov, Trump, Anklam, *et al.*, 2018; Dixon *et al.*, 2022), security (Palmer, Fukuyama and Relman, 2015)startup dependency (Freemont, 2019; Nylund *et al.*, 2022) in terms of achieving a steady stream of new innovation in the domain.

Social dynamics such as differing notions and rationales surrounding bioethics (Szocik *et al.*, 2021; Bohua *et al.*, 2023), social acceptance (Bauer and Bogner, 2020; Frow, 2020), also play a part.

, , , , , , .

## 4.3. Whack-a-mole governance

The most reasonable way to look at it would be: what can generative AI do within the frame of all of these barriers? From this we can wonder whether generative AI-enabled synthetic biology really *would* be truly different from previous approaches. We could, of course, also wonder how different the situation would be if many of those previously mentioned barriers somehow went away. Interestingly, what AI fanatics would respond is that once those barriers are gone, AIs would themselves produce such approaches that are far superior to what could be conceived by human experts. Alternatively, it is always possible that emerging, superior and multi-modal AI systems would be able to overcome enough barriers to transform the field anyway.

For now, the most prudent governance path seems to be to keep fostering a responsibility mindset in a distributed manner at global scale. Machine learning enabled digital processing is already improving diagnostic accuracy and reducing turnaround time for even complex lab tests (Undru *et al.*, 2022). The smart laboratory, with AI-automation of biosecurity, has arguably moved from concept to reality, enabling self-control process management flows of personnel, materials, water, and air, automated operation, automated risk identification and alarms (Li *et al.*, 2022). However, when technologies merge, uncertainties multiply (O'Brien and Nelson, 2020), cybervulnerabilities and circumvention options increase (O'Brien and Nelson, 2020). If it indeed was the case that AI lifts all boats, it wouldt mean that mediocre labs can more rapidly gain the ambition to modify their facilities and work practices, and start doing work regulated by BSL-3 and BSL-4 designations. But while more lab researchers than before might potentially deploy AI to carry out experiments that should be carried out in a lab with a stricter designation (a higher BSL), this would, in most cases, be against the regulations. Having said that, in India, for example, there are no national reference standards, guidelines, or accreditation agencies for biosafety labs, so those labs that do comply, rely on the international ones (Mourya *et al.*, 2014, 2017). China, also, lacks a comprehensive regulatory system for BSL-2 labs, and lacks trained biosafety staff (Wu, 2019). The numbers game is indeed instructional. The International Laboratory Accreditation Cooperation (ILAC) accredits over 88,000 laboratories (ILAC, 2023). If you consider that there are currently 64 BSL-4 labs, just imagine if all BSL-3 labs wanted to do BSL-4-type work, and were capable of it. There are already some 57 BSL-3+ labs (Kaiser,

2023). There are currently great experiments going on regarding the feasibility of rapid response mobile BSL-2 lab deployment to areas with a public health crisis, but those labs carry additional risks from rogue elements (Qasmi *et al.*, 2023). Or, what about if all BSL-2 labs (which include most labs that work with agents associated with human diseases) suddenly started doing BSL-3 or BSL-4 type work?  With AI, and without national control regimes, more BSL-2 labs will be tempted to think they can take on more advanced work, too quickly.

As has been pointed out, there is a need to deploy a governance continuum (Hamlyn, 2022). Based on our reworking of the issues based on the literature review (Linkov, Trump, Poinsatte-Jones, *et al.*, 2018), there are six governance levels (global, national, corporate, lab, scientist, citizens) and four governance types (precautionary, stewardship, bottom-up, and laissez-faire) to be considered in an emerging *Framework for AI-enabled synbio governance* (see Table X). Each of these need constant monitoring and renewal based on assessing threats, hazards, and opportunities. Each governance level might prioritize one approach, but must have aspects of all governance types. Each governance type must be reflected at all governance levels. Today, we only have elements of such a framework implemented, and the skills required to make a comprehensive approach happen are formidable and require an all-of-society effort.

At any level, the process is quite complex. For example, the corporate AI governance at British biopharma AstraZeneca includes compliance documents, a responsible AI playbook and consultancy service, an AI resolution board, and AI audits–emphasizing procedural regularity and transparency–and interlinking with existing procedures, structures, tools, and methods (Mökander *et al.*, 2022).

However, the literature review points to the fact that the true governance challenge is not only about the individual elements doing things right. Rather, proper governance is interactive, and adaptive, and requires working on all levels of governance simultaneously while not ignoring any one level for much time at all. One could describe the process as whack-a-mole governance (see Fig. 6) where there are many actors using small rubber mallets (aka laws, rules, norms, votes) that need to hit each level simultaneously for the button (risk) to stay down.

*Figure 6 Synbio's whack-a-mole governance challenge*

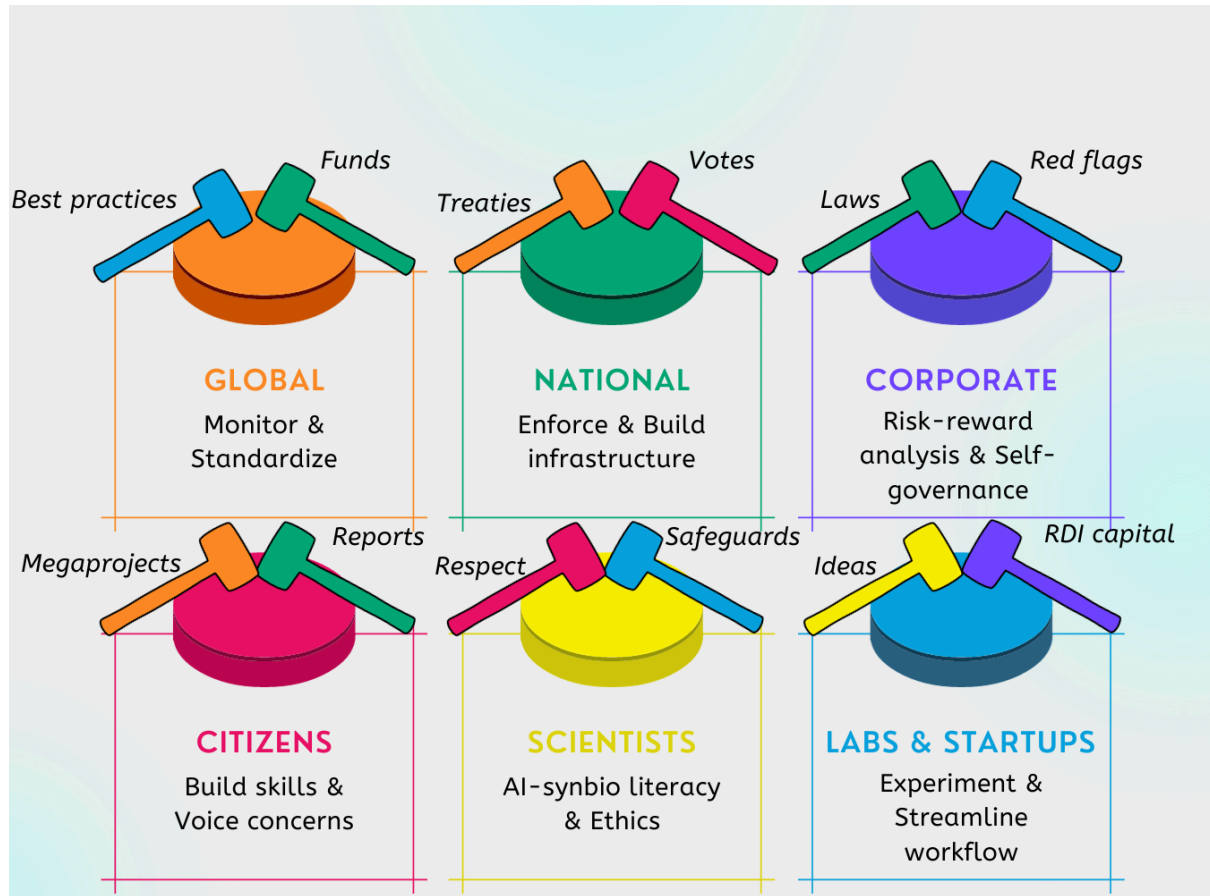

The above schematic must be complemented by transparent approaches for each set of tools (real-time monitoring, certification, compliance documentation, standardization, prizes, rewards). The fact that some types may adversely affect others, for example soft law might delay or undermine regulation or hard law, must be monitored and dealt with through responsible innovation (RI) approaches (Hamlyn, 2022). The entire governance structure (See Table 2) must work in a holistic way.

*Table 2. Framework for AI-enabled synbio governance*

| Level/Type | **Precautionary (command-and-control, hard law)** | **Stewardship (soft law)** | **Bottom-up** | **Laissez-faire (industry-driven)** |
|---|---|---|---|---|
| *Global* | Certification (BSL-framework), Precautionary principle, Intergenerational justice. GMO vs. SynBio agents | Standardization, Policy, Best practice sharing, Discussions, Recommendations | Adaptive, Global observatory model, Decentralization | Stakeholder discussions |
| *National* | Central authority<br>*USA*: Dual-Use Research of Concern (DURC), BMBL<br>*Israel:* DURC, Risk assessment, Xenobiology biocontainment systems w/genetic firewalls, Sustainability, Biodiversity | Innovation governance, R&D, Funding schemes, Subsidies Standardization, Surveillance, Indicator, Monitoring, Environmental surveillance (wastewater), Real-time data | Produce an evidence base, Stakeholder engagement, Regulatory flexibility, Value chain stimuli | Market making, Competition policy, Deregulation, Supply side policies |
| *Corporate* | Compliance documents, Responsible AI playbook, AI consultancy service, AI resolution board, AI audits, Screening DNA orders, International Gene Synthesis Consortium (IGSC) | Standardization (fora/consortia), company ethos, Experimental safeguards (competition, pathogenicity, predation, susceptibility, toxicity, allergenicity) | Adaptive (risk/reward), Timing of intervention, R&D biosafety checks, Trust, Investments | Lobbyism, deregulation push, self-governance, Risk–benefit analysis, Timing |
| *Labs & Startups* | Case-by-case, Approvals, Denials, Red flags (w/spot checks), Risk assessment, Safety-by-design (SbD) | Rules, Norms, Professional certifications, Indicators (Escape frequency, Strain fitness), Screening, Genetic safeguards, | Adaptive, Periodic review, Timing of intervention | Innovation, Product quality, Teams |
| *Scientist & Networks* | iGEM's safety and security programme, International Common Mechanism (by NTI), Precautionary principle | Rules, Norms, Ethics, Professional certifications, Screening, Flagging | Adaptive, Prizes, Periodic review, Responsibility, Timing of intervention (before work) | Self-regulatory |
| *Citizen* | Skills, News, Information campaigns | Norms, Ethics, Social acceptance | Adaptive, Skills, Awareness, participatory governance, Trust | DIY BIO, Citizen science labs, Participation, Voice, Risk perception |

# 5 Conclusion

The research question was: what are the most important emergent best practices on governing the risks and opportunities of AI-enabled synthetic biology. Indeed, some best practices are emerging, but it is still a disjointed picture. The first hypothesis that [1] there is a nascent literature on the impact of AI-enabled synthetic biology only found partial support. In fact the literature is nascent but there is scarce evidence on whether generative AI makes a big difference, or only adds to the emergence, and we had to consult related literature on generative AI in science and research to get closer to an answer.

The second hypothesis found more support, because [2] active stewardship is emerging as a best practice on governing the risks and opportunities of AI-enabled synthetic biology. Having said that, top-down governance, especially the command-and-control flavor, is not sufficient, and the literature points to decentralized governance as a remedy. A whack-a-mole governance model was formulated to describe and possibly also to address these challenges.

Hypothesis three which said that [3] to achieve proper governance, most, if not all AI-development needs to immediately be considered within the Dual Use Research of Concern (DURC) regime has some support. The larger point is that in some ways all research is dual use (Evans, 2022) because research always has many meanings and uses and compliance with the letter of imperfect, imprecise and rapidly outdated laws can only get you so far and also limits research in undesirable ways. Whose security are we trying to protect? Whose security typically is not protected? The DURC regime itself, instigated with the Fink report in 2003, is in serious need of an update in light of generative AI-enabled synthetic biology, and dual use is understood differently internationally (Lev, 2019). The review should begin immediately, but clarity on the threat is not likely to emerge for a few years, as generative AI-ready synbio-datasets and related functionality still needs to mature.

The last hypothesis [4] that even with the appropriate checks and balances, with AI-enabled synthetic biology, industrial biomanufacturing can conceivably scale up beyond the microscale within a decade or so, found some support but the field is still largely dependent on innovations that still have not materialized such as bioprocessing workflow, standardization of multi-omic datasets, and a design-build-test cycle that would be required to enable such scale-up. In fact, delving into the impact of AI-enabled synthetic biology for industrial biomanufacturing is a fruitful direction for future research.

At the end of the day it is safe to assume that AI-enabled synthetic biology is both a catalyst for risk (through creating novel synthetic organisms) and a potential for risk reduction and mitigation (through optimizing or restoring natural organisms and detecting pathogens). Governance of the phenomenon, and any attempts to megascale the bioeconomy in short order (by the US, UK, EU, China, or others) needs to keep both perspectives firmly in mind.

In closing, the premise of the article was that it is possible to identify best practices for governance, innovation, research, or policy on AI-enabled synthetic biology, and that these issues have commonalities and are best explored together. The subtext was to be more resilient towards risks but still being able to capture the opportunities. The topics do seem related, and relatable, but it is complex both for researchers, entrepreneurs, corporations, and policymakers to do so because of the transdisciplinary efforts required (Lee and George, 2023; Taylor *et al.*, 2023). However, in light of the revolutionary potential of AI-enabled synthetic biology, admittedly not yet fulfilled beyond single-cell microorganisms, one would have to conclude that best practices will change rather rapidly. If so, one implication might be that we chase such best practices in vain and that synthetic biology cannot deliver them (Hanson and Lorenzo, 2023).

Whack-a-mole type games seemingly are about quick reactions. However, it turns out that, according to the inventor of the version of the game with air cylinders, Aaron Fetcher, the best way to get a high score is to gaze in a relaxed way at the center of the playing field with the side moles in your peripheral vision (Brown, Fenske and Neporent, 2011). It is exactly that mix of calm focus with minimum effective effort which is needed for safe and sound AI-enabled synthetic biology scale-up. We are dealing with an environment with many possible distractions. As soon as one problem is fixed, another one will appear. Terminological and sectoral confusion, and growing pains within the industry, in the scientific establishment, and across the industries that are touched, will persist for some time. The obvious transdisciplinary barriers to growth are not easily or quickly resolved, even with a major reskilling effort. Generative AI might be a gamechanger, but biology will still be complex and surprising even to experts (and certainly surprising to machines). That's why emerging frameworks for AI-enabled synbio governance likely should contain a mix of precautionary (command-and-control, hard law), stewardship (soft law), bottom-up, and laissez-faire (industry-driven) approaches.

# Declaration of Generative AI and AI-assisted technologies in the writing process

During the preparation of this work the author did not use AI in the writing process. The author takes full responsibility for the content of the publication.

# Conflict of interests

Competing interests: The author(s) declare none.

# Funding

The study was partially supported by Open Philanthropy.

# Data availability

No data was used for the research described in the article.

# CRediT authorship contribution statement

**Trond Arne Undheim**: Conceptualization, Methodology, Data curation, Writing – original draft, Visualization, Investigation, Writing – review & editing.

**Tables**

*Table 1 List of biorisk standards and guidelines*

Global:
World Health Organization's (WHO) Laboratory Biosafety Manual (2004)
International Health Regulations (WHO 2005)
Biosafety in Microbiological and Biomedical Laboratories (BMBL), 6th ed. (Centres for Disease Control and Prevention (1984-2021) (BMBL, 2023)
Convention on Biological Diversity (CBD) and Protocols (Cartagena & Nagoya)
Biological and Toxin Weapons Convention (BTWC)
Tianjin Biosecurity Guidelines for the Code of Conduct for Scientists (2015)
World Economic Forum Global Future Council on Synthetic Biology

US:
International Compilation of Human Research Standards (ICHRS)
Laboratory Biorisk Management (CWA 15793)
Screening Framework Guidance for Providers of Synthetic Double-stranded DNA (U.S. HHS)
NIH Guidelines for Research Involving Recombinant or Synthetic Nucleic Acid Molecules
Executive Order 14081 to enable the progress of biomanufacturing and biotechnology (2022)
U.S. Executive Order on Safe, Secure, and Trustworthy Artificial Intelligence (2023)

EU:
Advanced Therapy Medicinal Products Regulation (ATMP) 2007

*Table 2. Framework for AI-enabled synbio governance*

| Level/Type | **Precautionary (command-and-control, hard law)** | **Stewardship (soft law)** | **Bottom-up** | **Laissez-faire (industry-driven)** |
|---|---|---|---|---|
| *Global* | Certification (BSL-framework), Precautionary principle, Intergenerational justice. GMO vs. SynBio agents | Standardization, Policy, Best practice sharing, Discussions, Recommendations | Adaptive, Global observatory model, Decentralization | Stakeholder discussions |
| *National* | Central authority *USA*: Dual-Use Research of Concern (DURC), BMBL *Israel:* DURC, Risk assessment, Xenobiology biocontainment systems w/genetic firewalls, Sustainability, Biodiversity | Innovation governance, R&D, Funding schemes, Subsidies Standardization, Surveillance, Indicator, Monitoring, Environmental surveillance (wastewater), Real-time data | Produce an evidence base, Stakeholder engagement, Regulatory flexibility, Value chain stimuli | Market making, Competition policy, Deregulation, Supply side policies |
| *Corporate* | Compliance documents, Responsible AI playbook, AI consultancy service, AI resolution board, AI audits, Screening DNA orders, International Gene Synthesis Consortium (IGSC) | Standardization (fora/consortia), company ethos, Experimental safeguards (competition, pathogenicity, predation, susceptibility, toxicity, allergenicity) | Adaptive (risk/reward), Timing of intervention, R&D biosafety checks, Trust, Investments | Lobbyism, deregulation push, self-governance, Risk–benefit analysis, Timing |
| *Labs & Startups* | Case-by-case, Approvals, Denials, Red flags (w/spot checks), Risk assessment, Safety-by-design (SbD) | Rules, Norms, Professional certifications, Indicators (Escape frequency, Strain fitness), Screening, Genetic safeguards, | Adaptive, Periodic review, Timing of intervention | Innovation, Product quality, Teams |
| *Scientist & Networks* | iGEM's safety and security programme, International Common Mechanism (by NTI), | Rules, Norms, Ethics, Professional certifications, | Adaptive, Prizes, Periodic review, | Self-regulatory |

|  | Precautionary principle | Screening, Flagging | Responsibility, Timing of intervention (before work) |  |
| --- | --- | --- | --- | --- |
| *Citizen* | Skills, News, Information campaigns | Norms, Ethics, Social acceptance | Adaptive, Skills, Awareness, participatory governance, Trust | DIY BIO, Citizen science labs, Participation, Voice, Risk perception |